\documentclass[fleqn,usenatbib]{mnras}

\usepackage[T1]{fontenc}
\usepackage{newtxtext, newtxmath}

\usepackage{graphicx}	
\usepackage{rotating}
\usepackage{multirow}

\usepackage{fancyhdr}
\title[Analysis of User Feedback]{Integrating Multi-Label Classification and Generative AI for Scalable Analysis of User Feedback}

\author[S. Loop et al.]{
Sandra Loop,$^{1}$\thanks{E-mail: sandra.loop@sap.com}
Erik Bertram,$^{1,2}$
Sebastian Juhl,$^{1,3}$ and
Martin Schrepp$^{1}$
\\
$^{1}$SAP SE, Dietmar-Hopp-Allee 16, 69190 Walldorf, Germany\\
$^{2}$Hochschule Fresenius Heidelberg, Sickingenstra\ss{}e 63-65, 69126 Heidelberg, Germany\\
$^{3}$University of Missouri, Columbia, USA
}

\date{Version: \today}

\pubyear{2026}

\begin{document}
\label{firstpage}
\pagerange{\pageref{firstpage}--\pageref{lastpage}}
\maketitle

\begin{abstract}
In highly competitive software markets, user experience (UX) evaluation is crucial for ensuring software quality and fostering long-term product success. Such UX evaluations typically combine quantitative metrics from standardized questionnaires with qualitative feedback collected through open-ended questions. While open-ended feedback offers valuable insights for improvement and helps explain quantitative results, analyzing large volumes of user comments is challenging and time-consuming. In this paper, we present techniques developed during a long-term UX measurement project at a major software company to efficiently process and interpret extensive volumes of user comments. To provide a high-level overview of the collected comments, we employ a supervised machine learning approach that assigns meaningful, pre-defined topic labels to each comment. Additionally, we demonstrate how generative AI (GenAI) can be leveraged to create concise and informative summaries of user feedback, facilitating effective communication of findings to the organization and especially upper management. Finally, we investigate whether the sentiment expressed in user comments can serve as an indicator for overall product satisfaction. Our results show that sentiment analysis alone does not reliably reflect user satisfaction. Instead, product satisfaction needs to be assessed explicitly in surveys to measure the user's perception of the product.
\end{abstract}

\begin{keywords}
User Comments, AI Supported Comment Classification, Sentiment Analysis, AI Supported Text Analysis, SBERT, Generated Comment Summaries
\end{keywords}

\renewcommand{\headrulewidth}{0pt}
\pagestyle{fancy}
\fancyhead{}
\fancyhead[LE]{\Large \thepage\hspace{.5cm} \textit{S. Loop et al.}}
\fancyhead[RO]{\Large \textit{Analysis of User Feedback} \hspace{.5cm} \thepage}
\fancyfoot{}

\thispagestyle{empty}

\section{Introduction}
User experience (UX) evaluation plays a central role in ensuring software quality, supporting informed design decisions, and fostering long-term product success. In highly competitive software markets, products that fail to meet users' expectations in terms of usability, usefulness, and overall experience are unlikely to achieve sustained adoption. Consequently, continuous UX measurement has become a core element of modern quality assurance and product management processes, particularly in large-scale and long-lived software systems \citep{brooke1996sus,schrepp_2017,lewis_sauro_2021,lewis_sauro_2021b}.

In practice, UX evaluation typically combines quantitative metrics derived from standardized questionnaires with qualitative feedback collected through open-ended questions. Well-established instruments such as the System Usability Score (SUS) \citep{brooke1996sus}, UMUX-Lite and UX-Lite \citep{lewis2013,lewis_sauro_2021}, UEQ-S \citep{schrepp_2017}, or the Net Promoter Score (NPS) \citep{reichheld2003one} provide compact, reliable, and interpretable numerical indicators of overall usability, user satisfaction, and perceived quality. These metrics are especially valuable for benchmarking, trend monitoring, and cross-product comparisons.

However, quantitative UX metrics alone are often insufficient to fully explain why users perceive a product in a certain way. For this reason, open-text user comments are frequently included in UX surveys. Such comments can provide concrete hints about specific strengths and weaknesses of a system, describe contextual factors that influence user perceptions, and highlight issues that are difficult to capture with predefined questionnaire items. In some cases, comments even point directly to actionable problems that can be transferred into concrete development tasks, such as bug fixes or usability improvements. As a result, the combination of quantitative metrics and qualitative comments is widely regarded as a best practice in UX evaluation.

Despite their potential value, user comments are notoriously difficult to analyze at scale. In continuous UX measurement programs, organizations may collect thousands of comments over time, making manual analysis costly and inefficient. Moreover, comments are often vaguely formulated, focus disproportionately on negative aspects, and vary greatly in length and specificity. A further challenge is the frequently observed mismatch between quantitative metrics and comment sentiment collected from the same respondent. For example, users may report high satisfaction or usability scores while simultaneously describing isolated negative issues in their textual feedback, resulting in an overall negative sentiment classification of the comment.

In our previous work, we addressed these challenges in the context of a long-term, large-scale UX measurement initiative and proposed methods to systematically transfer user feedback into product improvements \citep{hcii2025_Bertram}. Building on this foundation, the present paper focuses specifically on the analysis and interpretation of user comments and their relationship to quantitative UX metrics. Compared to quantitative measures obtained from standard questionnaires described above, the analysis of comments requires substantially more sophisticated methodological support.

In this paper, we describe techniques developed and refined within our long-term UX measurement project \citep{hcii2025_Bertram} to handle large volumes of user comments. These techniques include automated multi-label classification of comments into meaningful topic categories, the generation of structured summaries using generative AI (GenAI), and sentiment analysis. Furthermore, we present empirical analyses that compare the sentiment of user comments with responses to quantitative UX metrics derived from questionnaires. The goal of these analyses is to support UX professionals in interpreting comment sentiments appropriately and to avoid common pitfalls, such as equating negative comment sentiment with overall user dissatisfaction.

Overall, this work aims to demonstrate how qualitative and quantitative UX data can be combined in a complementary way: quantitative metrics provide robust indicators of overall experience quality, while user comments enrich these metrics with contextualized explanations, concrete examples, and actionable insights, provided that suitable analytical methods are applied.

\section{Developing a Multi-Label Comment Classification Model}\label{classification}
Open-text survey comments span from broad product impressions to fine-grained issue reports, making manual synthesis slow and inconsistent. Automated multi-label classification organizes this diversity into topic categories. Classifying comments based on their content also allows product teams to monitor changes in the share of comments related to a specific topic over time. For example, they can quickly verify if investments into performance improvements of a system lead, as expected, to a decrease in the share of comments that complain about slow response times.

Additionally, assigning each comment to topic categories also helps to search for user feedback related to a certain topic of interest. A product manager, for example, may be interested in identifying technical issues and errors users are facing in order to identify and address the underlying problems. Being able to filter out comments where users report system errors or bugs substantially reduces the time to identify and eventually resolve these issues.

In order to support product teams to efficiently use open-ended feedback from users, we develop and train a multi-label comment classification model using supervised machine learning and a human-in-the-loop approach that automatically assigns one or multiple topic labels to each comment. Our approach consists of multiple steps which we outline here.

\subsection{Deriving Topic Labels}
Since our supervised learning approach requires a set of pre-defined labels to train the model, we start by defining a topic taxonomy. The labels need to meet certain criteria for our use case. First, they need to be collectively exhaustive, implying that the labels are general enough to be broadly applicable across a diverse product portfolio and to be able to apply to user comments in the past as well as to those in the future. Second, they need to be consistently defined to ensure a high inter- and also intracoder reliability which is important for the quality of the manually labeled training dataset and, eventually, the performance of the classification model \citep{oconnor_et_al_2020}. Third, they need to represent topics relevant to the users, meaning that each label should not occur too infrequently (say in $>1\%$ of comments). Finally, the number of unique labels should be manageable such that they provide a quick and comprehensive overview.

In order to obtain labels that fulfill these criteria, we analyzed numerous sample comments from different products and at different points in time. We also conducted interviews with our target audience to ensure that we provide meaningful labels that help them in their work. After compiling an initial codebook with labels and their definitions, we reiterated the process to validate and refine the derived labels \citep{lichtenstein_2023}.

\subsection{Creating a Training Dataset}
To train our multi-label classifier, we created a corpus of user comments annotated with a validated topic taxonomy. The validated codebook comprises ten distinct topic labels.\footnote{Note that initially, we did not label user comments that were written in any language other than English but instead labeled them as 'Non-English' which constitutes the $11^{th}$ label. Out of the $2,756$ initially labeled comments, $215$ ($7.8\%$) were written in some language other than English. Since our second model update, we changed the process to labeling the translated comments using the ten topic labels.} Using this codebook, two human coders independently annotated $2,756$ user comments sampled from $30$ products and services on the SAP Business Technology Platform (BTP) between January 2022 and November 2023.

\begin{table}
  \centering
\begin{tabular}{|l|r|r|}
\hline
 & Count of & Share of \\
Label & Comments & Comments\\
\hline
Usability & 721 & 26.16\% \\
Functionality & 606 & 21.99\% \\
Error & 420 & 15.24\% \\
Other & 352 & 12.77\% \\
Performance & 339 & 12.30\% \\
General Feedback & 239 & 8.67\% \\
Help & 207 & 7.51\% \\
Visual Design & 114 & 4.14\% \\
Integration & 113 & 4.10\% \\
Licensing & 91 & 3.30\% \\
\hline
\end{tabular}
\caption{Count and share of comments per topic label in the initial training dataset.}
  \label{tab:topiclabels}
\end{table}

Since users regularly refer to multiple topics within a single comment, it is common to have multiple labels assigned to a single comment. In our initial training data, $20.86\%$ of comments have multiple labels and there are 169 unique label combinations of which the majority (89) only occur once. Table \ref{tab:topiclabels} also illustrates that users most frequently comment on the usability of the software as well as features and functionalities of a product.

\subsection{Generating Comment Embeddings}
Before training the classification model, we need to convert each comment into a vector of numbers called ``embeddings'' that the machine learning model can process. In order to preserve the semantic meaning, we generate word embeddings using an extension of the open-source \textit{fastText} library \citep{joulin2017} that contains subword information \citep{bojanowski2017}. Specifically, we use a pre-trained shallow neural network model with $300$ hidden nodes to obtain word vectors \citep{grave2018}. Next, we fine-tuned the model using $7,948$ user comments received from January 2022 until November 2023 in order to adapt the model to the more technical and SAP specific lingo of our users. Before generating word embeddings with the AI model, we perform typical text pre-processing steps, including the removal of stopwords, URLs, line breaks, tabs, and punctuation. We also lemmatize the comments to convert each word to its base form. The model provides word vectors (which are aggregated character $n$-grams \citep{bojanowski2017}). To obtain comment-level vectors, we average over the L2-normalized word vectors.

While the \textit{fastText} model handles out-of-vocabulary words very well due to the subword information, is computationally efficient, and makes fine-tuning very easy, embedding models based on a Transformer architecture generate contextual representations in which the embedding vector of a word varies with the context it is used in. This context awareness also implies that most text pre-processing steps should be avoided since they allow the model to learn word meanings from the context. More recently, we therefore explored the benefits of the Transformer architecture in our use case by implementing the sentence-based Bidirectional Encoder Representations from Transformers (SBERT) model \citep{devlin2019}. Specifically, we vectorized the comments in the training data using the fine-tuned \textit{fastText} model (including text pre-processing) and the SBERT model. The embeddings were then used as features in our classification model described below. The results show that generating embeddings using the SBERT model improves the micro-averaged F1 score by $24.58\%$. This improvement is especially notable as both model alternatives are using the same amount of data, showing the superiority of the Transformer architecture in this case.

\subsection{Specifying a Multi-Label Classification Model}
The embeddings generated from the user comments serve as features for our machine learning classification model. As shown above, users regularly mention different topics in a single comment, so the model needs to be capable of assigning multiple labels to a single comment. Furthermore, since many label combinations are unique, we cannot train a model on label combinations. Rather, we sequentially train binary classifiers for each label which allows the model to assign either no label, only one label, or many different labels to a comment. Another advantage of this approach is that we can investigate the performance of the model for each category separately.

After comparing the performance of different binary classifiers popular in data science such as multi-label logistic regression, k-nearest neighbors, support vector machines, and random forest, we found that, across all topic labels, the extreme gradient boosting algorithm performs best in terms of micro-averaged F1 score. It is an efficient implementation of the gradient boosting technique that has been proven to be very powerful in several different classification tasks \citep{friedman2001,friedman2000}.

\subsection{Hyperparameter Tuning}
Compared to other ensemble methods such as random forest models, extreme gradient boosting has several hyperparameters that need to be optimized to maximize its performance. To this end, we implement multi-label stratified $k$-fold cross-validation over a pre-defined parameter grid in order to identify the optimal hyperparameter setting that maximizes the micro-averaged F1 score. The parameter search space includes different settings of parameters such as the learning rate, minimum loss reduction, L1 and L2 regularization on the weights, or the maximum depth of individual trees. While the computational burden of this procedure notably increases with the parameter space, the thorough selection of hyperparameters is crucial for prediction quality.

\subsection{Incremental Learning from Human Feedback}
After deployment, we establish a human-in-the-loop process to monitor and correct predictions of the multi-label classifier. The goal is to prevent the degradation of the model and to continuously improve its predictive performance.

Using the validated codebook with the labels and their definitions, two human coders continuously evaluate and correct the classification done by the machine learning model. Using batch updating, the model gets retrained based on the newly available manually labeled user comments. By now, the training dataset for the classification model contains $8,757$ manually labeled user comments from $40$ different software products and services on SAP BTP. Table \ref{tab:finalmodel} provides a detailed overview over the number of comments per category as well as the predictive performance of our final multi-label classification model in terms of per-label precision, recall, and the F1 score. Precision measures the share of correctly labeled comments from all comments the model predicts to belong to the respective label. Recall is the share of correctly labeled comments from all comments that actually have this label. Finally, the F1 score is simply the harmonic mean of precision and recall. In most applications, the F1 score is the preferred metric to assess the predictive performance of a model because it optimized both, precision and recall at the same time.

\begin{table}
\setlength{\tabcolsep}{4pt}
  \centering
\begin{tabular}{|l|r|r||r|r|r|}
\hline
 & Count of & Share of & & & \\
Label & Comments & Comments & Precision & Recall & F1 Score\\
\hline
Usability & 2,656 & 30.34\% & 0.69 & 0.60 & 0.64 \\
Functionality & 2,085 & 23.82\% & 0.66 & 0.57 & 0.61 \\
Error & 1,300 & 14.85\% & 0.74 & 0.65 & 0.69\\
General & \multirow{ 2}{*}{1,182} & \multirow{ 2}{*}{13.50\%} & \multirow{ 2}{*}{0.64} & \multirow{ 2}{*}{0.59} & \multirow{ 2}{*}{0.62}\\
Feedback & & & & &\\
Performance & 1,028 & 11.74\% & 0.82 & 0.72 & 0.76 \\
Help & 993 & 11.34\% & 0.66 & 0.54 & 0.59 \\
Other & 903 & 10.32\% & 0.61 & 0.42 & 0.49\\
Visual Design & 534 & 6.10\% & 0.69 & 0.46 & 0.56 \\
Integration & 482 & 5.51\% & 0.77 & 0.49 & 0.59 \\
Licensing & 324 & 3.70\% & 0.87 & 0.78 & 0.82 \\
\hline
\end{tabular}
 \caption{Topic label frequencies and performance metrics (precision, recall, and F1 score) of the multi-label classification model.}
  \label{tab:finalmodel}
\end{table}

Comparing the topic frequencies, we see that the relative share of comments per category remains largely stable compared to our initial training dataset (see Table \ref{tab:topiclabels}). As expected, we find some differences in terms of model performance across the topic categories. This can largely be explained by the variety of issues summarized under one topic label which results in a very diverse vocabulary. The pronounced heterogeneity in specific issues raised in comments related to the residual topic 'Other', for example, leads to the worst predictive performance across all evaluation metrics shown here compared to all other categories.

Overall, despite the fact that only $3.7\%$ of comments belong to the topic 'Licensing', the model has the best predictive performance with regards to this category, followed by 'Performance' with an F1 score of $0.76$.



\section{Generating Comment Summaries}
Managers and product owners are keenly interested in understanding user perceptions of their products. While quantitative metrics provide important insights, they fall short in capturing the nuanced views expressed through user comments. However, sifting through hundreds of comments is impractical for stakeholders. Thus, there is a need for an effective method to convey this qualitative feedback succinctly.

To address this challenge, we use GenAI to generate summaries of user comments collected over extended periods, such as a quarter or an entire year. These summaries are incorporated into quarterly and yearly reports and are also made available through a dashboard accessible to all employees. In the dashboard, all the comments are included. In the reports, a selection of comments is shown that support the comment summaries. 

\subsection{Insufficient and Inefficient Results}
In the process, we adapted and refined our approach to generating comment summaries. 

Initially, we crafted prompts for each product requesting a general summary
of all comments. This proved biased and inconsistent: they over-weighted negative statements, varied in style across runs, and often included unsupported statements. This is likely due to the inherent nature of voluntary feedback that often skews toward negative sentiment and any positive feedback is often not specific, merely stating `It's great' or ``I love the usability', for example, which gets bundled into a short positive phrase.

Next, attempting to separate summaries into positive and negative sentiments produced confusing results. The topics included in the summary were often presented in a chaotic order, in that the order of topics shown were different in the positive versus negative paragraphs. Also, summaries sometimes presented contradictory statements without adequate context. The summary could say 'many thought the product was user friendly' then later say 'many thought the product was not user friendly.' Although these can both be true, it does not convey actual sentiment distribution --- were opinions evenly split, or did one sentiment predominate? And given the chaotic order, the positive statements e.g. 'is user friendly' and the corresponding negative statements e.g. 'is not user friendly' were very far apart causing the reader to re-read earlier sections to make sense of the summary. And the positive section was often almost as long as the negative section even when the negative responses far outweigh the positive. 

Without clear guidance of how the summaries should be formulated, GenAI often had incorrect statements --- either putting too much weight on single comments or hallucinating responses. 

We tried validating each statement against source comments removing statements that could not be found and reordering for coherence, but this was very labor-intensive. In some cases, removing unsupported claims left too little content, forcing manual rewriting. In this case, GenAI was not helping us be more efficient.

\subsection{Our Best Approach}
Eventually upon the request of product teams, we aligned our summaries with predefined comment categories, which yielded improved results. Our refined process involves:

\begin{itemize}
\item Creating the instructional GenAI prompt. The instructional prompt tasks GenAI to review the comments and classifications, then to produce a summary for each category that meets a threshold for the minimum number of comments (currently set at four for products with low comment volume; a higher threshold is used for products with high comment volume). GenAI is also tasked with identifying which specific comments informed each summary, listing their IDs and verbatim content. We found that this not only enhances summary quality but also facilitates easier validation. This instructional portion of the prompt is consistent for all the products. We had previously explored including much detail in the instructional portion that included a description of the specific product, but ultimately, it did not improve the quality of the output.

\item Adding product-specific comments to the prompt. The common instructional section of the prompt was then followed by the product-specific comments that include the comment IDs, comment text, and their classifications. The comment classifications were those as discussed in Section \ref{classification}.

\item Excluding products with too few comments. Note that if the total number of comments is less than 20, we do not try to create a summary as it becomes difficult to create a fair summary when the likelihood of each comment is unique, yet the summary cannot include all the comments.

\item Validating summaries and select verbatim comments. We validate the GenAI produced summaries and select verbatim snippets from user comments in a coordinated workflow. We start by skimming both the summary and the supporting comments for obvious inaccuracies. This step is typically quick, as extracts are limited to relevant portions of longer comments. 

In our reports, we present summarized insights alongside selected verbatim snippets. When selecting the snippets, we maintain the distribution of positive and negative sentiment to reflect the actual distribution in the dataset. We prioritize comments that directly support the summary statements. Similarly, we also ensure that the summary, includes positive phrasing if there are positive extracts found. The ``General Feedback'' category often helps to preserve this balance and phrasing. It often has most of the positive comments, as positive comments often do not fit neatly into other established categories (e.g., expressions like ``It's great.'' or ``I love it.'').

Comment selection serves as a validation step. For each unique attribute in the summary, we skim the associated extracts to select the ones that apply. Because GenAI outputs only relevant segments, it is straightforward to select excerpts that are already formatted for inclusion. 

In the typical case, generating the summary, validating it, and selecting comments is fast. However, some cases require manual intervention. If we cannot find a supporting comment for a summary attribute or if multiple attributes are semantically overlapping, we remove or consolidate those elements from the summary. If a summary is too long for our reporting constraints, we trim lower-priority or weakly supported attributes. When GenAI truncates extracts too aggressively, we retrieve the full comment via comment ID and expand the quote.

For products with few comments (i.e., 20 - 50), we pay particular attention to representativeness and avoid overgeneralization, as the comments may be distinct and not fall into themes well.

\item Using the summaries and snippets. This general approach to creating summaries and selecting sample verbatim comments results in summaries that describe what the respondents think of the product with respect to the relevant comment categories. Within the summary, we preface each topic with the comment category. E.g., 'Usability: Some users commented that ... Help: Documentation is viewed as ...' To keep the summaries brief, we typically only include the top four categories based on comment volume in the summary. Sometimes there is a category with fewer responses that we include because the responses are homogeneous in the opinion. Functionality and usability are common comment categories and are typically included in each summary.
\end{itemize}

As a final step, we make the report with the comment summary and supporting comment extracts available for review by key product stakeholders before sending the report to a broader audience.

\section{Interpreting Sentiments}
Sentiment analysis or opinion mining is a technique of natural language processing (NLP) \citep{pang2002thumbs,turney2002thumbs}. We use sentiment analysis to identify the subjective tone or emotional expression of the open-text survey comments. A typical classification scheme applied is to categorize comments as positive, negative, or mixed. But more nuanced classifications are possible.

The sentiment is an important attribute to filter the potentially huge list of comments collected in a certain period. Product owners or developers of an application are typically most interested in negative comments, since these sometimes indicate potential improvements or fields of action.

However, sentiment is often (mis)used as a proxy for overall satisfaction---for example, by comparing the ratio of positive to negative comments. Because voluntary feedback is negativity-biased and comments often target specific issues, this ratio may not reflect respondents’ overall experience. We therefore analyze the relationship between comment-level sentiment and satisfaction scores from two surveys, highlighting when these measures align and when they diverge.

\subsection{A Survey Concerning the Quality of Tutorials}
This survey evaluates tutorials that help developers and system administrators to get started with the SAP BTP. For complex products like cloud platforms that allow to develop and deploy applications a purely intuitive usage is typically not possible. Users need to acquire basic concepts, even if the user interface of the product is well designed. Tutorials that illustrate important work flows and tasks in the system are important for a quick on-boarding of new users. Thus, it is important to constantly check user satisfaction of such tutorials.

A corresponding survey is available inside all BTP tutorials over a feedback link and contains the following questions:

\begin{itemize}
\item The tutorial covers a realistic use case.
\item The tutorial contains only relevant information.
\item I was satisfied with the duration of the tutorial.
\item In my opinion the tutorial was well-structured.
\item The tutorial motivates me to learn more about this topic.
\item How likely is it that you would recommend this tutorial to a friend or colleague?
\end{itemize}

The first five questions can be answered on an 11-point rating scale with the end points {\it Strongly Disagree} and {\it Strongly Agree}. The average score of these items is used to calculate a tutorial quality score. Answers are scored from 0 (worst) to 10 (best), thus the tutorial quality score ranges from 0 to 10. The last question corresponds to the Net Promoter Score (NPS) \citep{reichheld2003one}, a single item metric. It can be answered on an 11-point rating scale with the end points {\it Not at all likely} and {\it Extremely likely}. 
 
In addition, learners can leave a comment below these rating items in a text area. The corresponding question is {\it How could we improve this tutorial? What other topics would you like to see? If you have any additional feedback, please provide it here}. The sentiment of a comment is derived by the corresponding functionality in Qualtrics which is the survey tool we use to deploy the surveys. Comments are classified as {\it Positive}, {\it Mixed}, or {\it Negative}.

Thus, for the participants who enter a comment (all questions are optional) we can relate the sentiment to the corresponding scores of the NPS or the tutorial quality score. Overall, the willingness to enter comments is relatively low. Approximately 12\% (548 responses) of the respondents leave a comment. The average length of comments is 123 characters in this survey. The comment length differs massively between the sentiment categories (Positive: 35, Mixed: 253, Negative: 153).

For the NPS, respondents are classified into {\it Promoters} (responses 9 or 10), {\it Passives} (responses 7 or 8), and {\it Detractors} (responses 0 to 6) \citep{reichheld2003one}. Table \ref{tab:NPS_Sentiment} shows the distribution of responses to the NPS and sentiment classifications.

\begin{table}
  \centering
\begin{tabular}{|c|c|c|c|}
\hline
  & Detractors & Passives & Promoters \\
  \hline
Negative & 120 & 52 & 154 \\
Mixed & 4 & 7 & 27 \\
Positive & 5 & 13 & 157 \\
\hline
\end{tabular}
 \caption{Distribution of responses concerning sentiment and NPS classification. Respondents providing high NPS ratings (\textit{Promoters}) write nearly the same amount of negative and positive comments. Nearly all respondents providing low NPS ratings (\textit{Detractors}) show negative comments.}
  \label{tab:NPS_Sentiment}
\end{table}

Based on Table \ref{tab:NPS_Sentiment}, we conduct a significance test to see if there is a relationship between comment sentiment and the NPS category that is generalizable beyond the user sample at hand. A chi-squared test reveals a positive and statistically significant correlation between the NPS category and the sentiment of a user comment ($\chi_{df = 4}^2 = 98.11$, $p = 2.5e^{-20}$). Cramer's V equals $0.30$, with a bootstrapped $95\%$ confidence interval within $[0.26; 0.35]$, indicating a strong relationship.

Using Table \ref{tab:NPS_Sentiment}, we can also obtain conditional probabilities in order to further investigate this relationship. In general, the conditional probability of an event $B$ is happening given that we observe event $A$, expressed as $Pr(B|A)$, is a function of the joint probability of $A$ and $B$ and the unconditional probability of event A: $Pr(B|A) = \frac{Pr(A \cap B)}{Pr(A)}$. Using this formula, we find that the conditional probability of being a \textit{Promoter} despite having observed a negative comment is $Pr(Promoter | Negative) = 47.24\%$ (95\% confidence interval $[41.88\%; 52.66\%]$).

We further investigate this surprising finding using a one-tailed binomial significance test assessing whether the probability of being a \textit{Promoter} given a negative comment is smaller than the probability of not being a \textit{Promoter} (\textit{Detractor} of \textit{Passive}) given a negative comment, that is: $H_0: Pr(Promoter | Negative) < Pr(No Promoter | Negative)$. The significance test fails to reject the null hypothesis at conventional levels of statistical significance ($p = 0.854$). When observing a negative comment, the conditional probability of being a \textit{Detractor} is $Pr(Detractor | Negative) = 36.81\%$ (95\% confidence interval $[31.76\%; 42.17\%]$). Thus, it is even smaller than the conditional probability of being a promoter when observing a negative comment. Hence, we conclude that a negative comment does not indicate dissatisfaction with the tutorial.

For the responses with a positive sentiment we see with a few exceptions only \textit{Promoters}. Thus, a positive sentiment seems to be an indication for high satisfaction. Mixed comments are rare, but in this category also \textit{Promoters} dominate.

Next, we look at the relation between sentiment and the tutorial quality score calculated from the first five questions in the survey. As expected, the tutorial quality score differs per sentiment (standard deviation in parenthesis):

\begin{itemize}
\item Negative Sentiment: 7.25 (2.82)
\item Mixed Sentiment: 8.60 (1.87)
\item Positive Sentiment: 9.40 (1.24)
\end{itemize}

\begin{figure*}
\centering
\includegraphics[scale=1]{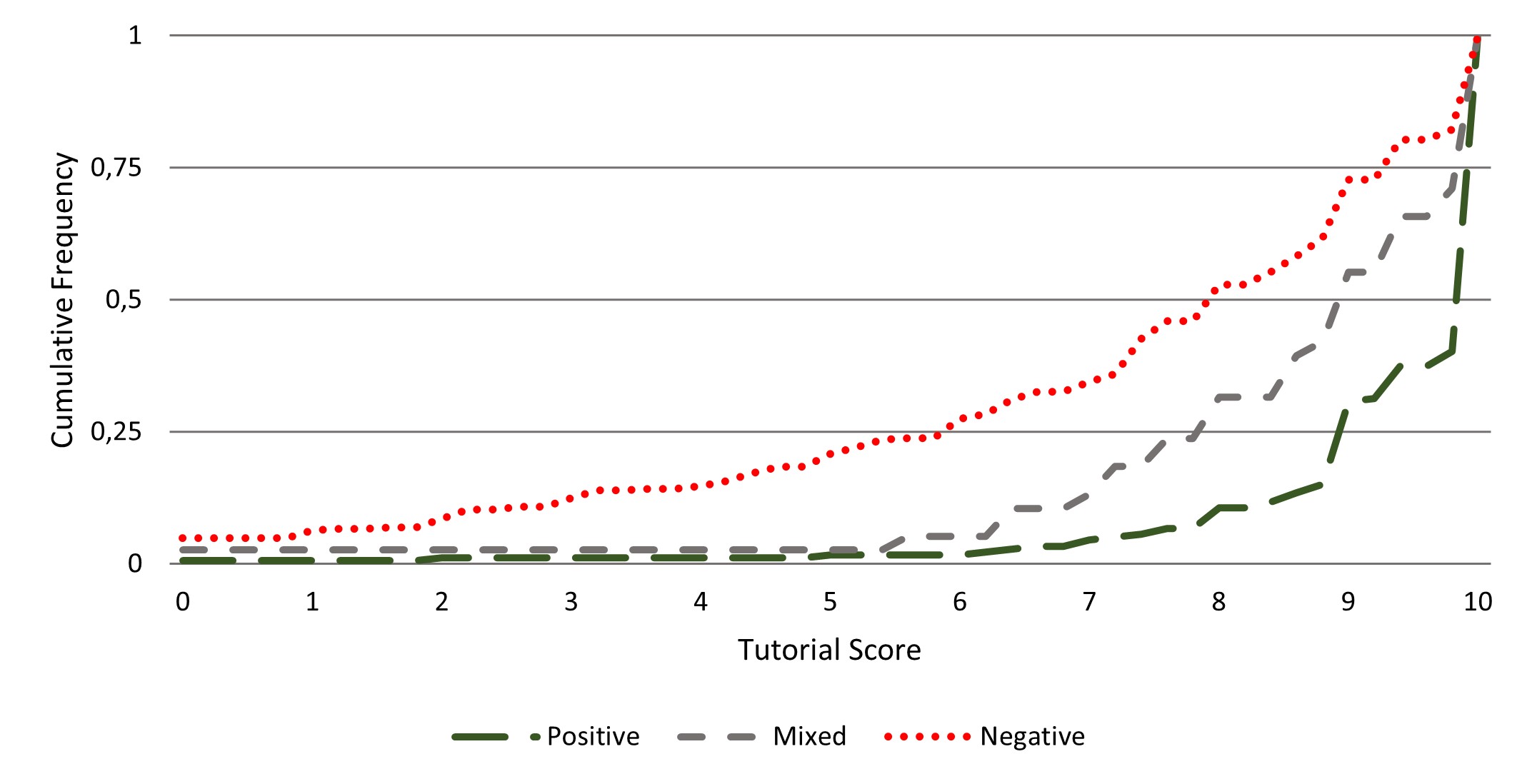}
\caption{Cumulative frequency of the tutorial quality score for the three sentiment categories. Respondents providing positive comments show with a few exceptions high tutorial quality scores. On the other hand, negative comments and high tutorial quality scores occur together quite frequently. Thus, a negative comment does not necessarily indicate a low satisfaction with the tutorial.}
\label{Fig:cumulative_frequency}
\end{figure*}

Even the group of participants with negative comments show a relatively high satisfaction with the tutorial quality (7.25). Figure \ref{Fig:cumulative_frequency} shows the cumulative frequency of the tutorial quality score for the three different sentiment categories.

We see again a picture similar to the NPS grouping. Respondents with a positive comment give high ratings concerning the score, i.e. show a high satisfaction with the tutorial. For example, 85\% of respondents with a positive sentiment show a score higher than 9. But even for respondents with a negative comment still 46\% show a score greater than 9. 

Overall, a positive sentiment of the comment implies a high satisfaction as measured by the tutorial quality score, while a negative sentiment does not imply a low satisfaction.

\subsection{A Survey Measuring the Usability of Applications}
This survey is available inside applications of the SAP BTP over a feedback link. The survey is quite short and opens as a dialog window. The goal is to collect feedback on the usability of applications from users and contains the following questions:

\begin{itemize}
\item Overall, how satisfied are you with \textit{<Application name>}.
\item \textit{<Application name>} does what it need to do.
\item \textit{<Application name>} is easy to use.
\item Please tell us about your experience with \textit{<Application name>}.
\end{itemize}

The first item corresponds to the common Product Satisfaction measure (PSAT) and can be answered on a 5-point rating scale, ranging from {\it Very dissatisfied} (1) to {\it Very satisfied} (5). The second and third question can be answered on a 5-point rating scale with the response options {\it Strongly disagree} (1), {\it Disagree} (2), {\it Neither agree nor disagree} (3), {\it Agree} (4) and {\it Strongly agree} (5). They correspond to the UX-Lite \citep{lewis_sauro_2021}. A score is calculated from the answers to these questions and results \citep{lewis_sauro_2021b,schrepp2023comparison} indicate that this score can be used as a reliable estimate of a SUS \citep{brooke1996sus} score, which allows to reuse the well-known SUS benchmark \citep{lewis_sauro_2018} for the UX-Lite. The last question refers to an open-text comment field and the sentiment of the entered comment is again derived by the built in functionality of Qualtrics. But for this survey the automatically assigned sentiments are manually checked and corrected if necessary by an expert. 

Again, the willingness to enter comments is quite low as only 7\% of respondents provide a comment. The average length of comments is 181 characters in this survey. The comment length differs massively between the sentiment categories (Positive: 53, Mixed: 212, Negative: 202). Table \ref{tab:Satisfaction} shows the distribution of responses to the different response options of the satisfaction question and the sentiment classifications.

\begin{table}
  \centering
\begin{tabular}{|c|c|c|c|c|c|}
\hline
  & Very & & Neither Satis. & & Very \\
  & Dissat. & Dissat. & nor Dissat. & Sat. & Sat. \\
\hline
Negative & 893 & 826 & 548 & 539 & 106 \\
Mixed & 45 & 66 & 161 & 345 & 135 \\
Positive & 35 & 3 & 15 & 232 & 380 \\
\hline
\end{tabular}
 \caption{Distribution of responses concerning satisfaction rating and sentiment classification. Nearly a quarter (22\%) of respondents with a negative comment still report satisfaction with the application. On the other hand, only a small number (6\%) of respondents with a positive comment report dissatisfaction.}
  \label{tab:Satisfaction}
\end{table}

Similar to the tutorial survey, a chi-squared test indicates a strong and statistically significant correlation between product satisfaction and comment sentiment ($\chi^2_{df = 8} = 1920.1$, $p \approx 0$).

We can see that a positive sentiment coincides in nearly all cases with a high satisfaction. In fact, the conditional probability of being very satisfied or satisfied with the product given a positive comment is $Pr(Very\ Satis. \lor Satis. | Positive) = 92\%$ with a $95\%$ confidence interval of $[89.72\%; 93.86\%]$. For the mixed sentiment a similar pattern emerges.

For the respondents who provided a negative comment, the picture is less clear. The conditional probability of being very satisfied or satisfied with the product given a negative comment is still $Pr(Very\ Satis. \lor Satis. | Negative) = 22.15\%$ with a 95\% confidence interval of $[20.68; 23.69]$. A one-tailed significance test shows that the difference in the probability of being satisfied or very satisfied and the probability of being very dissatisfied or dissatisfied given a negative comment is $Pr(Very\ Dissatis. \lor Dissatis. | Negative)  - Pr(Very\ Satis. \lor Satis. | Negative) = 36.88$ percentage points. While this difference is statistically significant ($p = \approx 0$), almost one quarter of respondents who provided a negative comment are still satisfied or very satisfied.

\begin{figure*}
\centering
\includegraphics[scale=1]{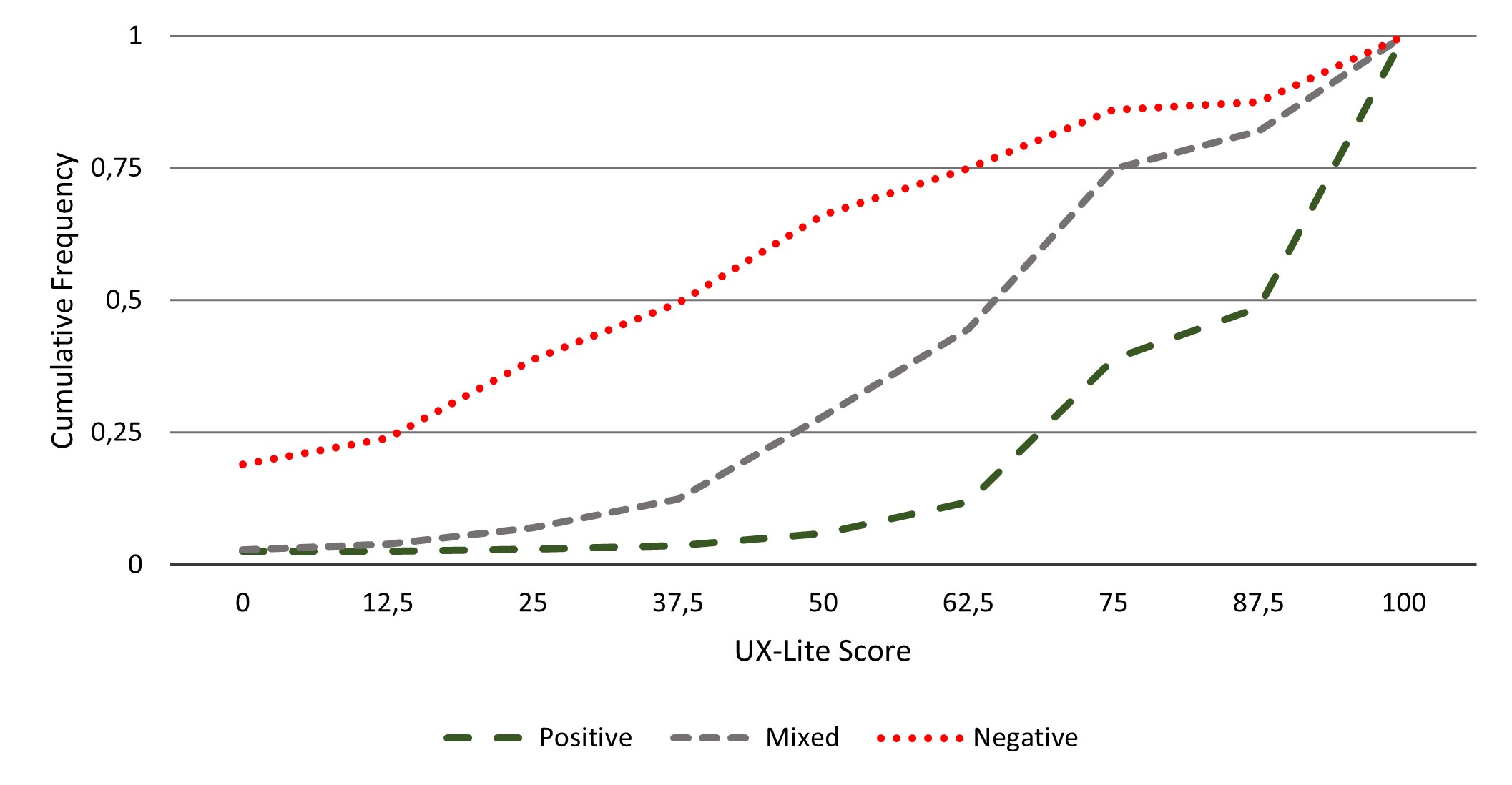}
\caption{Cumulative frequency of the UX-Lite scores for the three sentiment categories. Respondents that provide positive comments show with a few exceptions high UX-Lite scores. On the other hand, negative comments and high UX-Lite scores occur together quite frequently, i.e. negative comments do not imply a low overall satisfaction of users.}
\label{Fig:cumulative_UXLITE}
\end{figure*}

From the two questions corresponding to the UX-Lite a score between 0 and 100 can be calculated \citep{lewis_sauro_2021} that represents the usability of an application. Figure \ref{Fig:cumulative_UXLITE} shows the cumulative frequency of the KPI for the three different sentiment categories.

Similar to Figure \ref{Fig:cumulative_frequency} respondents with a positive comment give high UX-Lite ratings, i.e. show a high satisfaction with the usability of the application. And again we see that a substantial number of participants with a negative comment report high UX-Lite scores. 

\subsection{What Does This Mean for the Interpretation of Sentiments?}
We investigated two surveys concerning the dependency of the comment sentiments and metrics that measure the overall satisfaction of users. 

In both surveys we have seen that a positive sentiment is an indicator for a high satisfaction (measured by NPS and tutorial quality score in the first survey and by PSAT and UX-Lite in the second survey). 
But a negative sentiment is not a very good indicator for low satisfaction. Many promoters or users with a high rating concerning tutorial quality left a negative comment in the first survey. The same effect was found in the second survey concerning PSAT and UX-Lite. Thus, in conclusion a high ratio of negative sentiments should not be interpreted as an indicator for bad quality. 

How can we explain this result? First, the willingness of users to enter textual comments is relatively low in both surveys. In addition, the average length of positive comments is low compared to the average length of mixed and negative comments.

A manual investigation of the collected comments indicate that positive comments usually describe a general satisfaction with the tutorial or application without any details. On the other hand, negative comments describe in many cases detailed issues or error situations, for example difficulty to navigate to the desired part of the application, cases of slow system answers or detailed error codes in the system. 

Of course, such single negative events will have an impact on the general satisfaction, but this impact will be limited. It is unlikely that an isolated bad experience with a product will destroy a positive impression towards the product completely. But it is highly likely that such an experience will be mentioned in a feedback survey, since users hope that the issue will be fixed based on their comments.

We could see in both surveys that the scores indicating satisfaction (tutorial quality in the first survey and UX-Lite in the second survey) were significantly lower (but the total size of the effect is not too big) for participants that entered a comment compared to all participants who filled the survey. This also speaks for a relatively moderate effect of single bad experiences that are reported in the comments on the overall impression towards a tutorial or application.

As with all voluntary opt-in surveys, users self-select into the sample. Consequently, sampling bias becomes a concern since the data does not come from a random sample where all users have an equal probability of being sampled. While there is no optimal strategy to accurately quantify or even entirely remove the adverse impact of sampling bias on the UX KPIs, we apply post-stratification raking in order to improve the representativeness of the results for the overall user population. Details about this procedure and the application in our context can be found in \citet{hcii2025_Bertram}.

\section{Discussion \& Conclusion}
In this case study, we analyze end-user feedback using sentiment classification, topic classification, AI generated summaries, and sample supporting comments. We present the results in an interactive dashboard and in quarterly and yearly reports to make them accessible to a variety of stakeholders, allowing them to quickly get a wholistic understanding of what users think of the products.

The classification of comments based on their sentiments helps stakeholders to find comments with actionable insights; for example, negative or mixed comments can be reviewed for detailed feedback. The automated topic classification is useful to determine which aspects of the product users are concerned with the most. For instance, imagine that a product has a performance problem, so there are many comments about performance. As the product team works to improve performance, the expectation is to see fewer performance complaints. This would be reflected in the number of performance-related comments that are negative. Typically, product teams include survey data in their quarterly business reviews. They track whether they are improving against their KPI targets and include relevant comment analysis to provide context.

To avoid model drift and ensure a high quality of our classifications, we regularly update our topic classification model as products and markets evolve. For instance, after products started including AI features, we began receiving comments about AI. To account for this, we will add 'AI' as a new topic category in a future update of the model. Given the central role of AI in the business strategy, identifying these comments becomes crucial. Furthermore, we would like to provide an automated way to identify actionable comments. This indicator would be of great help for product teams as it would allow them to quickly identify issues they can immediately act upon. However, developing a model to classify comments based on this is challenging due to the difficulty in defining the target concept 'actionability'.

Based on the findings presented here, we derive practical recommendations for other teams interested in this approach. We recommend they define their own topic categories. As described above, we reviewed comments for our products to develop a set that applied to most products and was not too granular for the number of responses we received. Depending on the response volume, this approach can be easily extended to more granular topics. Some products may want to distinguish between comments about their product and comments about content created by their customers. However, in our case, there were not enough comments to accurately train the model to make this distinction.

When using GenAI to summarize comments, we recommend giving instructions that break down the tasks at hand. We did not find that providing additional information about who we are and why we need the data was productive. We also recommend establishing a process that quickly and easily validates the correctness of the summary. It is important to prevent decision-makers to make product decisions based on hallucinations or an overemphasis on a few responses.

Importantly, if a survey only has comments without a satisfaction or other UX metric, our results show that negative comments do not imply that users are unhappy with the product. They may already like it and have ideas for further improvements. Therefore, comment sentiments should not be viewed as an indicator of product satisfaction.

Besides these practical recommendations, our results suggest some directions for further UX research. Future research is needed that examines how different definitions of negative sentiment relate to UX metrics. Specifically, it remains to be investigated on how emotionally negative comments differ from non-emotionally comments with respect to their relationship to UX metrics. More generally, connecting the findings presented here to the literature on emotion detection and analytics \citep{nandwani2021} seems to be a promising way to gain further insights into end-users that could help product teams making the most of user comments.



\bibliographystyle{mnras}
\bibliography{bibliography}

@inproceedings{hcii2025_Bertram,
author = "Bertram, Erik and Hollender, Nina and Juhl, Sebastian and Loop, Sandra and Schrepp, Martin",
title = "{How to Transfer User Feedback into Product Improvements?}",
year = 2025,
pages={3--19},
doi = "https://doi.org/10.1007/978-3-031-93224-3_1",
booktitle = "M. Schrepp (Ed.): HCII 2025, LNCS 15795",
publisher = "Springer Nature Switzerland AG",
}

@inproceedings{lewis2013, 
  title={{UMUX-LITE: When there's no time for the SUS}}, 
  author={Lewis, James R and Utesch, Brian S and Maher, Deborah E}, 
  booktitle={Proceedings of the SIGCHI conference on human factors in computing systems}, 
  pages={2099--2102}, 
  year={2013} 
}

@article{reichheld2003one,
  title={The one number you need to grow},
  author={Reichheld, Frederick F},
  journal={Harvard business review},
  volume={81},
  number={12},
  pages={46--55},
  year={2003}
}

@article{schrepp_2017,  
  title={{Design and evaluation of a short version of the user experience questionnaire (UEQ-S)}},  
  author={Schrepp, M. and Hinderks, A. and Thomaschewski, J.},  
  journal={International Journal of Interactive Multimedia and Artificial Intelligence},  
  volume={4},  
  number={6},  
  pages={103--108},  
  year={2017}  
}

@article{brooke1996sus,
  title={{SUS: A quick and dirty usability scale}},
  author={Brooke, J},
  journal={Usability Evaluation in Industry},
  year={1996}
}

@inproceedings{pang2002thumbs,
author = {Pang, Bo and Lee, Lillian and Vaithyanathan, Shivakumar},
title = {Thumbs up? Sentiment classification using machine learning techniques},
year = {2022},
pages={79--86},
booktitle = {Proceedings of the ACL-02 conference on Empirical methods in natural language processing-Volume 10},
publisher = {Association for Computational Linguistics"}
}

@inproceedings{turney2002thumbs,
author = {Turney, Peter D},
title = {{Thumbs up or thumbs down? Semantic orientation applied to unsupervised classification of reviews}},
year = {2002},
pages={417--424},
booktitle = {Proceedings of the 40th Annual Meeting of the Association for Computational Linguistics},
publisher = {Association for Computational Linguistics,
Philadelphia, Pennsylvania, USA}
}

@article{schrepp2023comparison,
  title={{A Comparison of SUS, UMUX-LITE, and UEQ-S}},
  author={Schrepp, Martin and Kollmorgen, Jessica and Thomaschewski, J{\"o}rg},
  journal={Journal of User Experience},
  volume={18},
  number={2},
  year={2023}
}

@article{lewis_sauro_2018,
  title={{Item benchmarks for the System Usability Scale}},
  author={Lewis, J.R. and Sauro, J.},
  journal={Journal of User Experience},
  volume={13},
  number={3},
  year={2018}
}

@misc{lewis_sauro_2021,
  author    = {Lewis, J. R. and Sauro, J.},
  title     = {Measuring {UX}: From the {UMUX-Lite} to the {UX-Lite}},
  year      = {2021},
  howpublished = {\url{https://measuringu.com/from-umux-lite-to-ux-lite}},
  note         = {Accessed: May 19, 2022}
}

@misc{lewis_sauro_2021b,
  author       = {Lewis, J. R. and Sauro, J.},
  title        = {How to Estimate {SUS} Using the {UX-Lite}},
  year         = {2021},
  howpublished = {\url{https://measuringu.com/how-to-estimate-sus-with-ux-lite/}},
  note         = {Accessed: May 19, 2022}
}

@article{oconnor_et_al_2020,
author = {Cliodhna O'Connor and Helene Joffe},
title ={Intercoder Reliability in Qualitative Research: Debates and Practical Guidelines},
journal = {International Journal of Qualitative Methods},
volume = {19},
number = {},
year = {2020},
doi = {10.1177/1609406919899220},
}

@article{lichtenstein_2023,
author = {Matty Lichtenstein and Zawadi Rucks-Ahidiana},
title ={Contextual Text Coding: A Mixed-methods Approach for Large-scale Textual Data},
journal = {Sociological Methods \& Research},
volume = {52},
number = {2},
pages = {606-641},
year = {2023},
}

@inproceedings{joulin2017,
    title = "Bag of Tricks for Efficient Text Classification",
    author = "Joulin, Armand  and Grave, Edouard  and Bojanowski, Piotr  and Mikolov, Tomas",
    editor = "Lapata, Mirella  and Blunsom, Phil  and oller, Alexander",
    booktitle = "Proceedings of the 15th Conference of the {E}uropean Chapter of the Association for Computational Linguistics: Volume 2, Short Papers",
    year = "2017",
    address = "Valencia, Spain",
    publisher = "Association for Computational Linguistics",
    url = "https://aclanthology.org/E17-2068/",
    pages = "427-431"
    }

@inproceedings{grave2018,
  title={Learning Word Vectors for 157 Languages},
  author={Grave, Edouard and Bojanowski, Piotr and Gupta, Prakhar and Joulin, Armand and Mikolov, Tomas},
  booktitle={Proceedings of the International Conference on Language Resources and Evaluation (LREC 2018)},
  year={2018}
}

@article{bojanowski2017,
    author = {Bojanowski, Piotr and Grave, Edouard and Joulin, Armand and Mikolov, Tomas},
    title = {Enriching Word Vectors with Subword Information},
    journal = {Transactions of the Association for Computational Linguistics},
    volume = {5},
    pages = {135-146},
    year = {2017},
    month = {06},
    issn = {2307-387X},
    doi = {10.1162/tacl_a_00051}
}

@inproceedings{devlin2019,
    title = "{BERT}: Pre-training of Deep Bidirectional Transformers for Language Understanding",
    author = "Devlin, Jacob  and
      Chang, Ming-Wei  and
      Lee, Kenton  and
      Toutanova, Kristina",
    editor = "Burstein, Jill  and
      Doran, Christy  and
      Solorio, Thamar",
    booktitle = "Proceedings of the 2019 Conference of the North {A}merican Chapter of the Association for Computational Linguistics: Human Language Technologies, Volume 1 (Long and Short Papers)",
    month = jun,
    year = "2019",
    address = "Minneapolis, Minnesota",
    publisher = "Association for Computational Linguistics",
    doi = "10.18653/v1/N19-1423",
    pages = "4171-4186",
}

@article{friedman2001,
 author = {Jerome H. Friedman},
 journal = {The Annals of Statistics},
 number = {5},
 pages = {1189--1232},
 publisher = {Institute of Mathematical Statistics},
 title = {Greedy Function Approximation: A Gradient Boosting Machine},
 urldate = {2024-12-02},
 volume = {29},
 year = {2001}
}

@article{friedman2000,
 author = {Jerome Friedman and Trevor Hastie and Robert Tibshirani},
 journal = {The Annals of Statistics},
 number = {2},
 pages = {337--374},
 publisher = {Institute of Mathematical Statistics},
 title = {Additive Logistic Regression: A Statistical View of Boosting},
 urldate = {2024-12-02},
 volume = {28},
 year = {2000}
}

@article{nandwani2021,
  title={A review on sentiment analysis and emotion detection from text},
  author={Pansy Nandwani and Rupali Verma},
  journal={Social Network Analysis and Mining},
  volume={11},
  number={81},
  pages={1--19},
  year={2021}
}



\label{lastpage}
\end{document}